Attilio Sacripanti EJU Scientific Commission
**The increasing importance of Ashi Waza, in high level competition.**
*(Their Biomechanics, and small changes in the form).*

**Abstract**

In this paper we are interested, both: at the description of Biomechanics of the Ashi Waza techniques, with precision of few of them, members of the Japanese, so called, Ko Waza group; and at the biomechanical analysis of the small changes in the applicative form of them.
These small changes, from the basic movement defined by Kodokan and show in thousand books, probably born from practical competitive situations. They are today become the usual applicative form, not only present among the various National Federations, but also in Japan itself among Universities dojos and the Kodokan. Among the Ko Waza techniques, we will select for our analysis only the following: Ko Uchi Gari , Ko Soto Gari, Ko Uchi Barai, Ko Soto Barai, Ko Soto Gake , Ko Uchi Gake
That are considered by Japanese followers different throwing techniques, as the different names show us. But from the Biomechanical point of view all these Kodokan throws are based on the same physical principle and practically on the same movements.

1. *Introduction.*
2. *Updating Technical Information.*
3. *Basic Biomechanics of leg movement in Ko Uchi - Ko soto.*
4. *Minor Changes*
5. *Conclusions.*
6. *References.*

Attilio Sacripanti EJU Scientific Commission
**The increasing importance of Ashi Waza, in high level competition.**
*(Their Biomechanics, and small changes in the form).*

1. Introduction

In this paper we are interested, both: at the description of Biomechanics of the Ashi Waza techniques, with precision of few of them, members of the Japanese, so called, Ko Waza group; and at the biomechanical analysis of the small changes in the applicative form of them.

These small changes, from the basic movement defined by Kodokan and show in thousand books, probably born from practical competitive situations. They are today become the usual applicative form, not only present among the various National Federations, but also in Japan itself among Universities dojos and the Kodokan. Among the Ko Waza techniques, we will select for our analysis only the following: Ko Uchi Gari , Ko Soto Gari, Ko Uchi Barai, Ko Soto Barai, Ko Soto Gake , Ko Uchi Gake

That are considered by Japanese followers different throwing techniques, as the different names show us. But from the Biomechanical point of view all these Kodokan throws are based on the same physical principle and practically on the same movements.

The small changes, introduced in the judo practice in these years, have introduced small variations in the shape of the movement showed by Kodokan as perfect form of throwing, leaving unchanged the biomechanical principle that holds them.

The best representation in the books regarding the executive form of these techniques, we find it without a shadow of doubt in the golden text of Kazuzo Kudo 9th Dan *[4]* the last Kano's student, who in the Ashi Waza of our group mentions and shows the following throwing techniques: Ko Uchi Gari, Kouchi Barai , Kouchi Gake,  Kosoto Gari,  Kosoto Barai, Kosoto Gake,  Nidan Kosoto Gari, Nidan Kosoto Gake.

That you can see in order in the next eight figures:

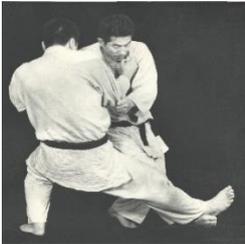 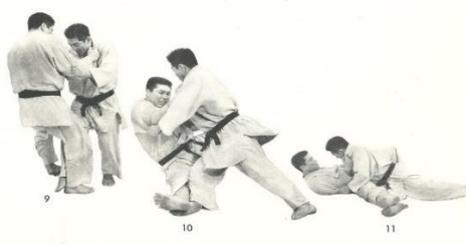 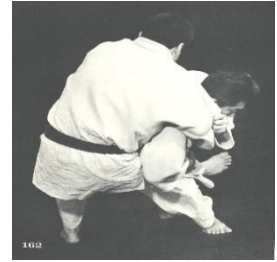

*Fig.1-3  Ko Uchi Gari,            Ko Uchi Barai                                              Ko Uchi Gake*

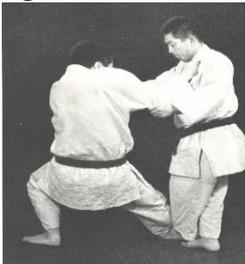 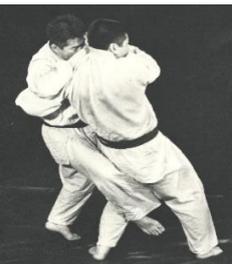 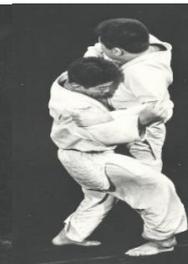 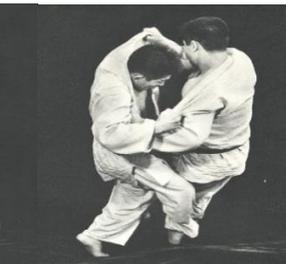 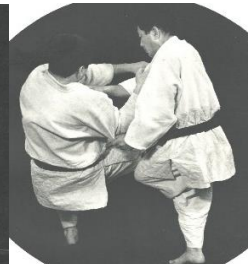

*Fig.4-8 Kosoto Gari ,   Kosoto Barai      Kosoto Gake,      Ni Dan Kosoto Gari,       Ni Dan Kosoto Gake*

One of the best analysis of Ko Uchi  Gari history name is presented in  *[5]*  in which the authors affirm that *"Ko-uchi-gari [minor inner reaping throw] is a versatile throw of which the exact origin is unknown"*. Again, it is possible to read in this very interesting paper "*Clear and correct jūdō-specific and didactic terminology are helpful in acquiring the complex motor skills that enable a refined and highly technical jūdō". [5].*

The *"Didactic terminology"* used by Japanese Masters is very useful to help the proficient judo students to understand and categorize the complex movements that build the judo throwing techniques, but there are some shadows in it. The problem, with the Japanese terminology of Kodokan techniques, is that it is often of generalist type thus impeding to highlight the deep biomechanical connections, that exist in them and thus preventing an approach of global methodology of teaching. Despite the different names used to characterize what "for the Japanese Masters" may appear as different techniques, it is easily to see that all these "different" techniques are based on the same mechanical movement, to hook, to reap and / or to sweep Uke's foot with Tori's foot. In term of biomechanics there is no differences among them, regarding the applied physical principle, the only thing different is the kind of resistance that the swept foot shows, for low friction the sweeping action is sufficient, for high friction to reap or to hook is better, this means that force must increase from Barai action to Gari or Gake.

## 2. Updating Technical information

In these last times, in high level competition, the use of Ashi Waza is increased, probably due to the new referee regulations. See Tab 1,2,3,4, in both functions as winning techniques and /or most used.

| Rio Olympic 2016   Winning Nage Waza  (Female) | | |
|---|---|---|
| **Te Waza** | 15 | 20% |
| **Ashi Waza** | 32 | 42%   (**25%**)* |
| **Koshi Waza** | 16 | 22% |
| **Ma sutemi** | 1 | 0,01% |
| **Yoko Sutemi** | 12 | 16% |
| | | |
| **Lever** | 44 | 58% |
| **Couple** | 32 | 42% |

**Tab 1  Winning techniques at Olympic Rio 2016   ( Female)** *[1] (*Personal Elaboration)

| Rio Olympic 2016 Winning Nage Waza ( Male) | | |
|---|---|---|
| **Te Waza** | 42 | 27% |
| **Ashi Waza** | 55 | 35%  (**24%**)* |
| **Koshi Waza** | 29 | 19% |
| **Ma sutemi** | 16 | 10% |
| **Yoko Sutemi** | 14 | 9% |
| | | |
| **Lever** | 101 | 65% |
| **Couple** | 55 | 35% |

**Tab 2  Winning Techniques at Olympic Rio 2016   ( Male)** *[1] (*Personal Elaboration)

| Rio Olympic 2016 Winning Nage Waza  (Total) | | |
|---|---|---|
| **Te Waza** | 57 | 26% |
| **Ashi Waza** | 87 | 37%  (**24%**)* |
| **Koshi Waza** | 45 | 19% |
| **Ma sutemi** | 17 | 7% |
| **Yoko Sutemi** | 26 | 11% |
| | | |
| **Lever** | 145 | 63% |
| **Couple** | 87 | 37% |

**Tab.3 Winning Techniques at Olympic Rio 2016  ( Total)** *[1]* (Personal Elaboration)

| World Championship 2017  <span style="color:red">Most used</span>  Nage Waza | | |
|---|---|---|
| **Te Waza** | 152 | 34% |
| **Ashi Waza** | 226 | 50%  (**62%**)* |
| **Koshi Waza** | 31 | 7% |
| **Ma sutemi** | 21 | 5% |
| **Yoko Sutemi** | 19 | 4% |
| | | |
| **Lever** | 223 | 49,7% |
| **Couple** | 226 | 50,3% |

**Tab.3 <span style="color:red">Most Used</span> Techniques  2017 World Championship** *[2]* ( Personal Elaboration)

In the first three months of 2019 we have this trend as winning techniques application, (see tab 4 )

| (March) 2019 Winning Nage Waza | | |
|---|---|---|
| **Te Waza** | 68 | 26% |
| **Ashi Waza** | 120 | 46%  (**32%**)* |
| **Koshi Waza** | 26 | 10% |
| **Ma sutemi** | 14 | 5% |
| **Yoko Sutemi** | 36 | 13% |
| | | |
| **Lever** | 144 | 54% |
| **Couple** | 120 | 46% |

**Tab.4 Winning Techniques in the IJF competition in first Three Months of 2019.** *[3]*

*Percentage of winning Ko Waza techniques utilized among Ashi Waza.

## 3. Basic Biomechanics of leg movement in Ko Uchi - Ko soto

In term of Biomechanics the group of techniques described before in picture, can be seen as the same application of the **Couple** applied by Tori by Arms and one leg *[6 ]*, starting in two parallel planes to his transverse plane and continuing along the trajectory defined by the rotation of Uke's body around his center of Mass, ( the presence of the external gravity force produces noticeable variations in this trajectory and head and feet of Uke move along a path which differs from a perfect circumference, centered in its center of mass).  The action of the Tori's foot is simple is a movement that appear similar to the pendulum movement with a final collision against the Uke's foot. The force applied could be different if the Uke's foot is moving on the Tatami (low friction) sweeping action is necessary with only a low impact force, if the weight of Uke is on the foot well stable, ( high friction : first detachment ) hooking or reaping action are necessary with a high impact force. However, if we analyze closer the only one mechanic model that will describe these techniques; the leg action could be modeled like a physical pendulum who collides the adversary's foot.  Remembering the previous discussion about the friction variability connected to the so called actions: barai , gari, and gake, it is possible define a good modeling of these actions.

This model could be the model that describes a variable dumping physical pendulum, in connection with the mechanics of an anelastic collision between feet.*[6]* These models are well known both in Physics and Biomechanics. Remembering the anatomy of legs (fig.9 -11)

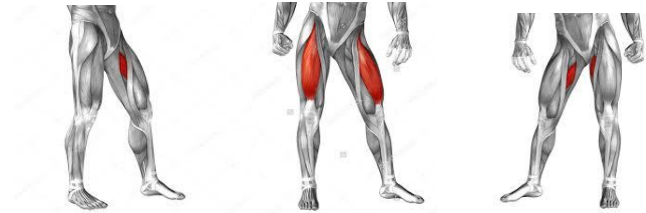

*Fig 9,10,11. Adductor muscles that work during Ko Uchi-Ko Soto adduction Movement*

Normally the inferior kinetic chain action could be considered a physical pendulum, a pseudo-rigid body that undergoes, in first simpler approximation, a fixed axis rotation about a fixed point.

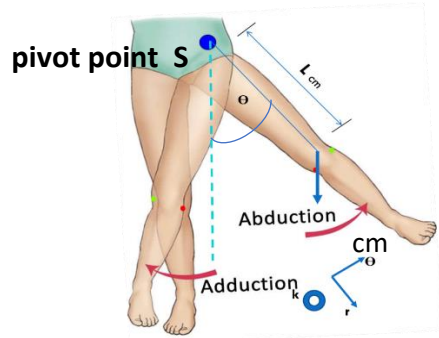

*Fig 12 Biomechanics of Ko soto-Ko uchi*

The torque T applied by the muscular structure about the pivot point *S* is given by the component of gravity force and the resultant of muscular action.

$$T = T_{cm} \wedge (mg + T_r) = l_{cm} r \wedge (mg + T) [\cos\theta \, r - \sin\theta \, \theta] \cong -l_{cm}(mg + T)\sin\theta \, \mathbf{k} \quad [1]$$

When θ > 0 the torque about *S* is directed in the negative ˆ**k** -direction (into the plane of page) when θ < 0 the torque about *S* is directed in the positive ˆ**k** -direction (out of the plane of page)

The moment of inertia of the center of mass about the pivot point *S* is $I_s$.

The rotational equation of motion is then, considering with good approximation, $T_r \approx -T\sin\theta$ [2]

$$-(T + mgl_{cm})\sin\theta = I_s \frac{d^2\theta}{dt^2} \qquad [3]$$

$$\frac{d^2\theta}{dt^2} - \frac{(mgl_{cm} - T)}{I_s}\sin\theta = 0 \qquad [4]$$

The addition of damping to Eq. [4] makes it analytically unsolvable.
Assuming that the damping is proportional to the angular velocity, the equation of motion becomes:

$$\frac{d^2\theta}{dt^2} + \mu\frac{d\theta}{dt} - \frac{(mgl_{cm} - T)}{I_s}\sin\theta = 0 \quad [5]$$

There are no chaotic solutions of Eq. [5].
Almost all solutions of this equation describe phase space trajectories terminating at the stable fixed point:
$\theta = \frac{d\theta}{dt} = 0$ which *"attracts"* all trajectories from its *"basin of attraction."*

Another important notation is that, the equation is a kind a Langevin rotational equation, and if the torque applied by muscular contraction is a random force, then the solution, it is well known, is a Rotational Brownian Motion on very short interval of time and microscopic space.

But as already demonstrated in *[7]* for the linear analogous, the same for the rotational ones, as predicted by the Hamilton equations of the athlete's system

$$\frac{d\theta}{dt} = \omega = \frac{L}{I_s} \quad [6] \qquad \frac{dL}{dt} = T - \mu L \quad [7] \qquad \text{The solutions are easily evaluated:}$$

$$L(t) = T\tau + (L_0 - T\tau)e^{-\frac{t}{\tau}} \qquad [8]$$

$$\theta(t) = \theta_0 + \frac{T\tau}{I_s}t + \frac{\tau}{I_s}(L_0 - T\tau)\left[(1 - e^{-\frac{t}{\tau}})\right] \qquad [9]$$

*the stationary solutions $t \gg \tau$ are*

$$L(t) \approx T\tau \qquad [10]$$

$$\theta(t) \approx \theta_0 + \frac{T\tau}{I_s}t \qquad [11]$$

As well known, to account for the sharp distribution in mechanical terms, Langevin added to the right-hand side of equation a stochastic force **ξ**(*t*), also called *white noise*. It is a random function of time with zero mean and a covariance proportional to the Kinetic energy produced, in our case." athletes fight", by the work of push-pull forces. For which: $\langle \xi(t) \rangle = 0$.

Working out Langevin's picture, it is found that if *f* is the solution of the so-called Klein–Kramers equation,

$$\frac{\partial f}{\partial t} + \frac{L}{I_s}\frac{\partial f}{\partial r} + \frac{\partial}{\partial L}\left[\left(T - \mu L - \frac{\mu I_s \omega^2}{2}\frac{\partial}{\partial L}\right)f\right] = 0. \quad [12]$$

Letting $E_k = \frac{\mu I_s \omega^2}{2} = 0$ in equation [12] gives a partial-differential equation on *f* which, mathematically speaking, admits the Hamilton equations [6] and [7] as its characteristics

Langevin's approach is cast in the language of rotational dynamics with a novelty, namely the stochastic force **ξ**(*t* ). When the latter is input into rotational Newton's second law of motion, we obtain a stochastic differential equation.

Langevin's equation looks more intuitive at one sight and indeed was put forth before, its mathematics is subtle and open to criticism.

Defining the average of any observable *O* from the probability density *f* in the usual way of classical statistical mechanics:

$$\langle O \rangle \equiv \iint Of(\theta, L, t)d\theta dL \qquad [13]$$

It is possible to show by algebraic manipulations that equation [13] entails

$$\frac{d\langle\theta\rangle}{dt} = \frac{\langle L\rangle}{I_s} \qquad [14] \equiv [6]$$

$$\frac{d\langle L\rangle}{dt} = T - \mu\langle L\rangle \qquad [15] \equiv [7]$$

Because in our situation T and μ are independent of position the Hamilton equations [7] and [8] are recovered on average.

Then the Newtonian approach is connected to the average on long time and space, while the Brownian motion characterizes very short time and microscopic space of observation

In the mechanical model of Ko Uchi- Ko Soto movement, it is not important the whole free motion which could be harmonic, but only the first part of it, till to the feet collision. Now we can consider the equation that drive the Barai Action. In such application, in the general equation [6] it is possible to consider the μ negligible (~ 0) and the equation [6] becomes equal to the equation [5].

Instead the equation [6] will better model the Gari- Gake action in function of the increasing friction between Uke's foot and the tatami, given by the amount of body weight that rests on it.

As it is possible to see in the next theoretical explicative figure,

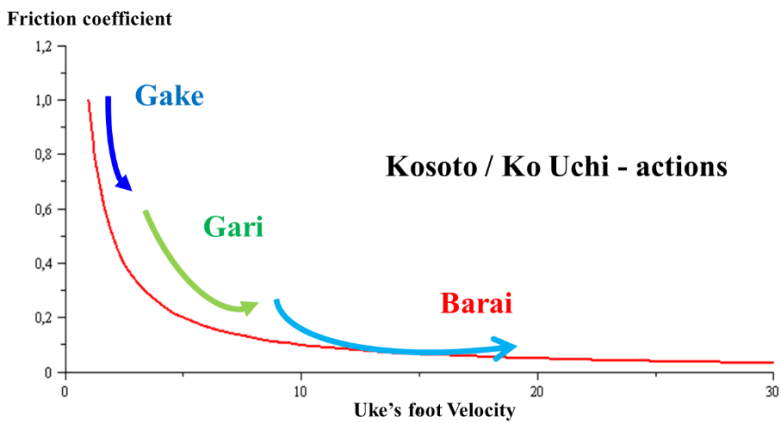

*Fig 13 Friction coefficient between Uke foot and tatami and Gake -Gari- Barai actions*

The impact force can be easily evaluated by the collision theory, considering an elastic impact for simplicity, in the impact point. If the impact of the feet is like a plastic-collision (feet connected to each other after the collision) we can write prior to the moment of collision between the foot with body of mass $M_1$ and the leg with body mass M, the Tori foot has a high tangential velocity V.

After the collision, the foot quickly binds to the other foot by lifting the body and imparts a velocity v to it. Linear momentum before and after the collision is preserved so that:

$$MV = (M_1 + M)v \qquad [16]$$

After the foot is connected with the other foot, the kinetic energy of the combination of feet thrusts the Uke's body outward and upward to a height h. All the kinetic energy is transformed to potential energy and Therefore:

$$\frac{1}{2}(M_1 + M)v^2 = (M_1 + M)gh \qquad [17]$$

Solution of Eqs. [16] and [17] yields the velocity of the Tori's foot.

$$V \approx \frac{(M + M_1)}{M}\sqrt{2gh} \qquad [18]$$

From this result we can evaluate the formula of the effective impact force, whose variability is linked to the contact speed divided by the square root of the double height

$$F = M\frac{dV}{dt} \cong (M + M_1)\sqrt{2g}\,\frac{d}{dt}\left(\sqrt{h}\right) \cong (M + M_1)\sqrt{2g}\,\frac{d}{dx}\left(\sqrt{h}\right)\frac{dx}{dt} \qquad [19]$$

Whit simple calculations we obtain

$$F = M\frac{dV}{dt} \cong (M + M_1)\sqrt{\frac{g}{2h}}\,v \qquad [20]$$

As numerical example we can evaluate the force produced by a leg of a 71Kg athlete, against a 69Kg adversary, if the whole leg weights 24Kg and the adversary's foot is raised to 7 cm, and the weight on the adversary's foot is approximately half of the total weight ( 34,5 Kg) and the impact velocity is v=0,8 m/s the force applied is about F≈390 N that is the force to accelerate 390 Kg at 1 m/s.

4. **Minor Changes**

Today as already previously indicated around the world and in the Japan too little changes are introduced into the Kodokan form, these variations probably born from competitive application are , for some people, they seemed more effective, or more suited to their morphological structure, so much so that in Japan for example Nakamura Yukimasa performs a Kosoto Gari very different from the Kodokan Form ( Fig 4) using the thigh to sweep, instead of the foot, this kind of variation is the type shown in the following figures.

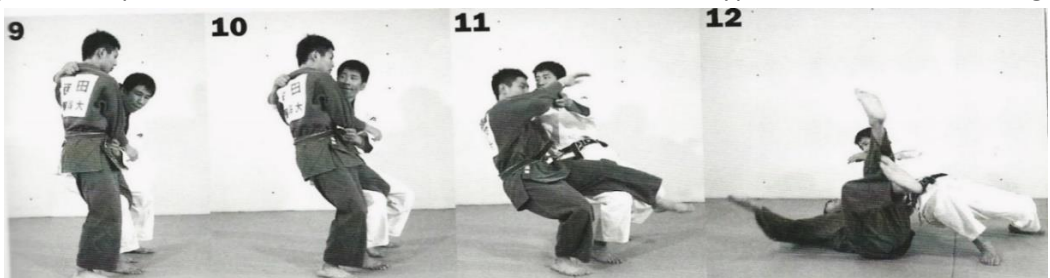

*Figg.14-17 Modern variation of Kosoto Gari [8]*

Same thing just a little bit different we can see in the execution performed by Tobitsuka Masatoshi

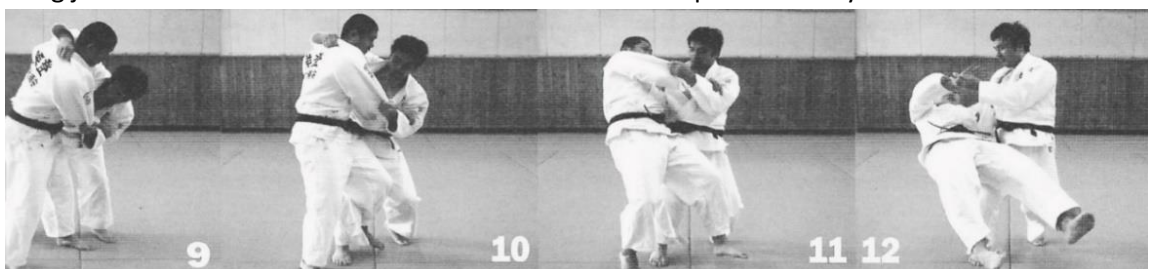

*Figg.18-21 Modern variation of Kosoto Gari [8]*

Who in the next sequence shows a different form of Kosoto Gari, in a quick tactical attack at the beginning of the fighting, very similar to the off-grips Korean Kosoto Gari attack, on the opportunity of a mistake of the opponent's grips.

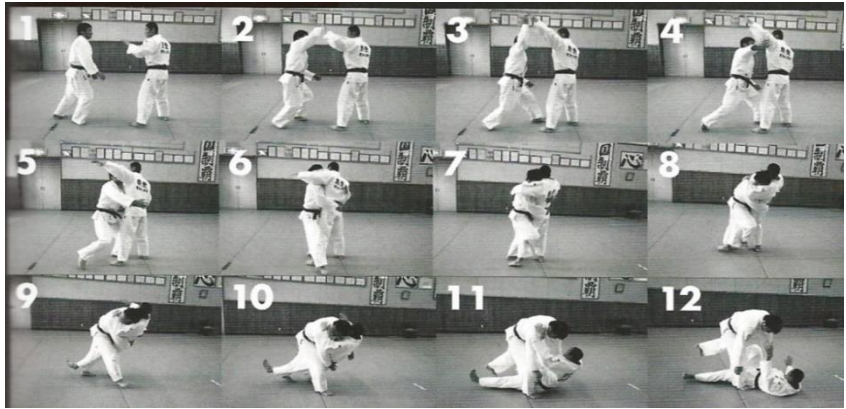

*Fig.22 Tactical variation of Kosoto Gari [8]*

In the next figures we can see the personal interpretation of Kosoto Gake of Masuchi Katsuyuki

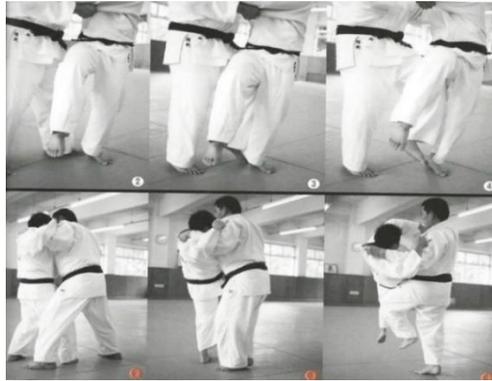

*Figg.23-28 Modern interpretation of Kosoto Gake [8]*

This kind of variations are a multipurpose application because if Uke's resistance is stronger, the technique can always be changed from a Couple Application into a lever variant by rotating Uke on the thigh.

Not only Japanese Champions give new personal interpretations of the classical form proposed by Kodokan, but contribution come also from other countries, very well-known are the Russian or the Korean contributions, ( Like Kabarelli or Reverse Seoi ). In our examination focused on the Ko Waza of the Couple's group, the next figures show a very effective French Interpretation of competitive Ko Uchi Gake, performed by Larose.

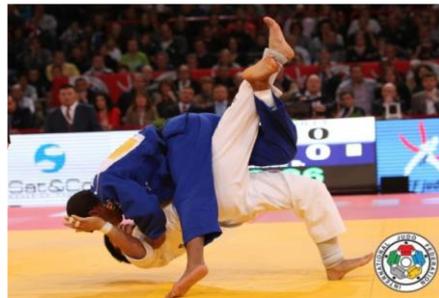

*Fig.29 Modern French competitive form of Ko Uchi Gake*

Less interpretative changes, have been seen around the world, about Ko Uchi Gari. It seems that the Kodokan form of this technique, could be almost perfect, in itself.

Ko-Uchi-Gari in the Biomechanical Classification is classified as a "Couple applied by Arm(s) and Leg". Ko-Uchi-Gari requires a simultaneous opposing set of forces (Couple) provided by the *Tori*'s arms driving the *Uke* to his rear corner and *Tori*'s foot sweeping *Uke*'s foot in the opposite direction, thus forwards, the pendular motion before analyzed. In Ko-Uchi-Gari, the forces operate in the Tori's transversal plane. Tori, who is applying, the "couple," thus moves in the sagittal plane. At the same time, the movement incorporates a rotation around the vertical axis involving the trunk/leg compartment which encompasses the coxo-femoral articulation. These two concomitant actions highlight the flexibility the Tori needs in order to perform Ko-Uchi-Gari effectively.  Important is, maintaining the control, not to bend the torso and carry out the action as straight as possible (pushing the *hara* toward the *tatami*), so as to apply the torque as simultaneously as possible. This is one of the most common mistakes. Control should not end at the beginning point of the reaping and Uke starting to fall, but should be maintained throughout the final point. In the next sequence Fukumi Tomoko shows a competitive and very effective application of classical Ko Uchi Gari, in which it is possible to found all the previous described phases .

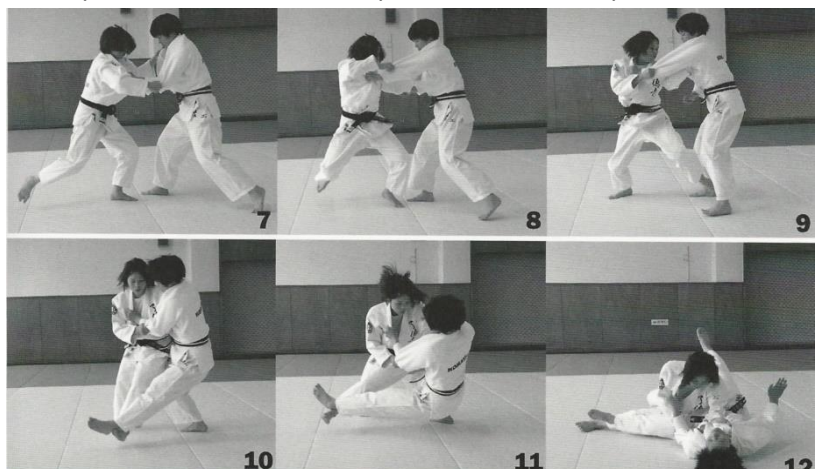

*Figg.30-35 Modern interpretation of Ko Uchi Gari [8]*

A different interpretation of Ko Uchi comes from Korea, for example the Korean An Cang uses it in combination with Suwari Seoi and applies on the imbalance deriving from the first technique a very wide Ko Uchi as a block of Uke's foot and not as a sweep, achieving victory.However, all these minor changes, if they change the classical form proposed by the Kodokan, they do not change its basic mechanics. All the interpretative variants seen, belong to the Couple group and are modeled by a pendulum model (actually three-dimensional and not one-dimensional as shown).

These small techniques today show a growing notoriety, that is often combined, not only with their effectiveness, but also with the brevity of the movement and its practical simplicity.

5.  Conclusion

In this short paper it is shown the biomechanical models of Ko Uchi, Ko Soto , throwing techniques, that are increasing in importance, because of their effectiveness in competition, so as to increase their presence as winning techniques in international competitions.

All these techniques are classified in the Couple group with a Couple applied by Arms, and Leg, as seen before the basic mechanics is the same the only difference is the increasing forces and Couple applied against the increasing resistance die to increasing friction, between tatami and Uke's feet.

It is significant that all new interpretive variants have been developed in techniques in which the actions of Gari and Gake are applied. This is certainly due to the need for increased strength and to the fact that the new variants or make the technique more flexible with regard to the increased defensive capabilities at international level, such as Japanese variants shown in Fig (14-28) or drastically decrease these capacities making the technique of fact indefensible, such as the French variant in Fig (29).

The action of Ko Uchi Gari has not undergone any major changes, which shows that its effectiveness is such that, even with the agonistic evolution, it retains all its effectiveness in the synthetic form identified by the Kodokan, the Korean interpretation changes the basic mechanics of the technique altering a Couple technique in a Lever with maximum arm.